\documentstyle[12pt]{article}    
\begin{document}

\vspace{2cm}

\begin{center} {\large \bf Fluxbrane in Melvin Background: Membrane Matrix Approach}
                                                  
\vspace{1cm}

                      Wung-Hong Huang\\
                       Department of Physics\\
                       National Cheng Kung University\\
                       Tainan,70101,Taiwan\\

\end{center}
\vspace{1cm}
\begin{center} {\large \bf  Abstract} \end{center}
We apply the regularized theory of light-cone gauge memberbranes to find the matrix model in the Kaluza-Klein Melvin background.   In the static system the Melvin matrix model becomes the BFSS model with two extra mass terms and  another term.   We solve the equation of motion for simplest N=2 D0 branes system and find that there is a nontrivial configuration.   Especially, we also make an ansatz to find the nonstatic solution.  The new solution shows that $D_0$  propagating in the magnetic tube field background may expand into a rotating noncommutative ring.

\vspace{1cm}
\begin{flushleft}
E-mail:  whhwung@mail.ncku.edu.tw\\
\end{flushleft}
\newpage
\section  {Introduction}
The Melvin metric [1,2] is a solution of  Einstein-Maxwell theory, which describes a static spacetime with a cylindrically symmetric magnetic flux tube.   It provides us with a curved space-time background in which the superstring theory can be solved exactly [3].   In the Kaluza-Klein spacetime the Melvin solution is a useful metric to investigate the decay of magnetic field [4,5] and the decay of spacetimes, which is related to the closed string tachyon condensation [6-8].   The fluxbranes in the Melvin spacetime have many interesting physical properties as investigated in the resent literatures [9,10].

   In this paper we try to investigate the branes dynamics in the Kaluza-Klein Melvin background from the membrane matrix model approach.   Historically, the Melvin matrix model had been discussed by Motl [11], while the Melvin matrix string model had also been discussed by Sekino and Yoneya [12].  However, to arrive our purpose we will first explicitly derive the Hamiltonian of  the Melvin matrix model.

   In section II we review the Melvin spacetime and make efforts to represent the metric in a suitable form.   The metric form is used to obtain a discreted theory of membranes in the Kaluza-Klein Melvin background, with the help of the general scheme of de Wit, Peeters and  Plefka [13].  After the quantization the Melvin matrix model is then obtained.    For any value of the twist parameter, i.e. external magnetic flux tube field is arbitrary,  the Melvin matrix model becomes a simple form of the BFSS model [14] with two extra mass terms and another term.   In section III we solve the equation of motion of the simplest  two D0 branes system and  have found a nontrivial solution.   The results show that there is a nontrivial configuration therein.  In section IV, we try to find the nonstatic solution by making an ansatz .  The new solution we found shows that two $D_0$  propagating in the magnetic tube field background may expand into a rotating noncommutative ring.   Last section is devoted to a discussion.

\section  {Hamiltonian of Melvin Matrix Model}
The Melvin spacetime is a solution of the Einstein-Maxwell theory [1].  One of the most surprising features of the Kaluza-Klein Melvin spacetime is that the corresponding higher dimensional spacetime is a flat manifold subject to non-trivial identifications [4,5].   Explicitly,  the 11-dimensional flat metric in M-theory is written in cylindrically coordinates
$$ds^2=-dt^2+ \sum_{m=1}^{7}dy_m dy^m +d\rho^2+\rho^2  d\varphi ^2 + dx_{11}^2 ,  \eqno{(2.1)} $$
with the identifications
   $$(t,y_m ,\rho,\varphi, x_{11}) \equiv (t,y_m,\rho,\varphi+2\pi n_1 R B +2\pi n_2, x_{11}+2\pi  n_1R), \eqno{(2.2)} $$
The identification under shifts of  $2\pi n_2$ for $\varphi$ and $2\pi  n_1R$ for $x_{11}$ are, of course, standard. The new feature is that under a shift of $x_{11}$, one also  shifts $\varphi$ by $2\pi n_1 R B$.  More geometrically, one can obtain this spacetime by starting with (2.1) and identifying points along the closed orbits of the Killing vector  $l=\partial_{11}+B\partial_\varphi$.  Therefore we can introduce 
the new coordinate $\tilde\varphi=\varphi-Bx_{11}$ which is constant along the orbits of $l$ and rewrite the metric as 
$$ds^2=-dt^2+ \sum_{m=1}^{7}dy_m dy^m +d\rho^2+\rho^2  (d\tilde\varphi+Bdx_{11}) ^2 + dx_{11}^2 ,  \eqno{(2.3)} $$
with the points $(t,y_m ,\rho,\tilde\varphi, x_{11})$ and $(t,y_m,\rho,\tilde\varphi+2\pi n_2, x_{11}+2\pi  n_1R)$ identified. 

    Since the eleven-dimensional spacetime is flat this metric is expected to be an exact solution of the M-theory including higher derivative terms. We can recast the eleven-dimensional metric in the following canonical form  [4,5]
$$ ds_{11}^2= e^{-2\phi/3}ds_{10}^2+  e^{4\phi/3} (dx_{11}+2 A_\mu dx^\mu )^2 , \eqno{(2.4)}  $$
the ten-dimensional IIA background is then described by
$$ ds_{10}^{2} = \Lambda^{1/2}\left(-dt^2+\sum_{m=1}^{7}dy_mdy^m+d\rho^2\right)
  +\Lambda^{-1/2}\rho^2d\tilde{\varphi}^2 ,\eqno{(2.5)} $$
$$  e^{4\phi/3}=\Lambda \equiv 1+\rho^2B^2 ,~~~~ A_{\tilde{\varphi}}=\frac{B\rho^2}{2\Lambda}.  \eqno{(2.6)} $$
The parameter $B$ is the magnetic field along the $z$-axis defined by
$B^2=\frac{1}{2}F_{\mu\nu}F^{\mu\nu}|_{\rho=0}$.  The Melvin spacetime is therefore an exact solution of M-theory and can be used to describe the string propagating in the magnetic tube field background.

    To proceed, we first define the light cone coordinates
$$x_\pm \equiv t \pm x_{11},\eqno{(2.7)} $$
and use the diffeomorphisms of the spacetime to define the new coordinates $(X_+, X_-,R,\Phi, Y_m)$ by the relations
$$X_+ \equiv x_+, ~~~ R \equiv \rho,~~~~~Y_m \equiv y_m,\eqno{(2.8)}$$
$$d\tilde\varphi \equiv {\sqrt{2}\over \sqrt{2+B^2}}d\Phi + {B\over \sqrt{2+B^2}}dX_-,\eqno{(2.9)}$$
$$dx_- \equiv  {-B\over \sqrt{2+B^2}}d\Phi + {\sqrt{2}\over \sqrt{2+B^2}}dX_-,\eqno{(2.10)}$$
In terms of the new coordinates the metric becomes

$$ds^2=-{2 \sqrt{2}\over \sqrt{2+B^2}}dX_+ dX_-  + {B^2(X^2+Y^2)\over 2}dX_+^2 + dX^2 +dY^2 + {B^2\over 2(X^2+Y^2)} (XdY-YdX)^2$$
$$+{2 B \over \sqrt{2+B^2}}\left(1+{1+{B^2/2}\over X^2+Y^2} \right) (XdY-YdX)dX_+ +  \sum_{m=1}^{7}dY_m dY^m, \eqno{(2.11)} $$
\\
in which we define $X\equiv R~cos(\Phi), Y\equiv R~sin(\Phi)$.  The metric tensor in the above new coordinate has the property $g_{--}=g_{a-} =0$, which is a necessary condition to apply the formula of [13] to find the matrix model in the Melvin spacetime.

  In reference [13] de Wit, Peeters and  Plefka had studied  the membrane Lagrangian density (we neglect the fermion parts in this paper)
$$L= -\sqrt {-g} -{1\over6}\epsilon^{ijk}\partial_i X^\mu\partial_j X^\nu\partial_k X^\lambda C_{\mu\nu\lambda}. \eqno{(2.12)} $$
Upon fixing the light-cone gauge, they find that the bosonic part of the total Hamiltonian becomes 
$$ H=\int d^2\sigma {\Big (} {G_{+-} \over P_- - C_-} {\Big [}
{1\over 2} (P_A - C_A - {P_- - C_- \over G_{+-}} G_{A+} )^2 + {1\over 4}
(\epsilon^{rs} \partial_r X^A \partial_s X^B)^2 {\Big ]} $$
$$- {P_- - C_- \over 2 G_{+-}}G_{++} - C_+ + {1\over P_- - C_-}
{\Big [} \epsilon^{rs} \partial_r X^A \partial_s X^B P_A C_{+-B} + 
C_- C_{+-} {\Big ]} {\Big )}, \eqno{(2.13)}$$
where 
$$  C_A = -\epsilon^{rs} \partial_r X^- \partial_s X^B C_{-AB} 
+ {1\over 2}\epsilon^{rs} \partial_r X^B \partial_s X^C C_{ABC},\eqno{(2.14)}$$
$$C_{\pm} = {1\over 2} \epsilon^{rs} \partial_r X^A \partial_s X^B 
C_{\pm AB},\hspace{4.5cm}\eqno{(2.15)}$$ 
$$C_{+-} = \epsilon^{rs} \partial_r X^- \partial_s X^A C_{+-A}\hspace{4.5cm}.\eqno{(2.16)}$$
In the Melvin background $C_{\mu\nu\lambda}=0$ and we can easily use the metric form (2.11) to find the light-cone Hamiltonian. 

$$ H = { \sqrt{2}\over \sqrt{2+B^2}}{\Bigg \{}{1\over2}\left[\dot X + {B\over \sqrt 2}\left(1+{1+B^2/2\over X^2+Y^2}\right)Y \right]^2 \left(1+{B^2 X^2 \over 2(X^2+Y^2)} \right) \left({2  \over 2 +B^2} \right)~~~~~~~~~~~~~~~~~~~~~~~~~~~~ $$

$$+ {1\over2}\left[\dot Y - {B\over \sqrt 2}\left(1+{1+ B^2/2\over X^2+Y^2}\right) X \right]^2 \left(1+{B^2 Y^2 \over 2(X^2+Y^2)} \right) \left({2  \over 2 +B^2} \right)~~~~~~~~~~~~~~~~~~~~~~~~~~$$

$$~~~~~~+ \left[\dot X + {B\over \sqrt 2}\left(1+{1+B^2/2\over X^2+Y^2}\right)Y \right]\left[\dot Y - {B\over \sqrt 2}\left(1+{1+ B^2/2\over X^2+Y^2}\right) X \right]\left({B^2 XY \over 2(X^2+Y^2)} \right) \left({1\over 2 +B^2} \right)$$

$$ +~{1\over4}~{\Big [}\epsilon^{rs} \partial_r X \partial_s Y {\Big ]}^2 \left(2+B^2+{B^4\over 4}\right) {\Bigg \}} -{\sqrt2 \sqrt {2+B^2} \over 8}B^2 \left( X^2+Y^2 \right) ~~~~~~~~~~~ \eqno(2.17)$$
\\
However, the Hamiltonian is to complex to be studied.   Therefore some approximations shall be used.     If we consider the static system then the Hamiltonian become

$$H_{static}={-\sqrt{2}B^2\over \ 8 (2+B^2)^{3/2}}\left[\left(4B^2+B^4\right)\left(X^2+Y^2\right) - {\left(2+B^2\right)^2\over X^2+Y^2}\right] - {1\over 16} [X,Y]^2 {\sqrt {2} \left[4+(2+B^2)^2\right] \over \  (2+B^2)^{1/2}},  \eqno(2.18)$$
\\
in which we have used the replacement  
$$ \epsilon^{rs} \partial_r X^A \partial_s X^B=\{ X^A,X^B\} \rightarrow -i [X^A, X^B], \eqno{(2.19)}$$
and used the scale $P_-=- 1$ [13].  Note that the terms which are those containing the $Y_m$ are irrelevant to our investigation and therefore are neglected. 

   The equation of motion associated with  Hamiltonian (2.18) is 

$$ - Y [X,Y] \left[4+(2+B^2)^2\right]-{B^2\over \  (2+B^2)}\left[\left(4B^2+B^4\right) + {\left(2+B^2\right)^2\over (X^2+Y^2)^2}\right] X = 0, \eqno{(2.20)}$$

$$  X [X,Y]  \left[4+(2+B^2)^2\right] -{B^2\over \  (2+B^2)}\left[\left(4B^2+B^4\right) + {\left(2+B^2\right)^2\over (X^2+Y^2)^2}\right] Y = 0. \eqno{(2.21)}$$
\\
In the next section we will try to find some solutions of the above equation.

\section  {Static Solutions of Melvin Matrix Model}
At first sight it seems difficult to solve the above equations as $X$ and $Y$ are the $N$ by $N$ matrix.   However, we will try to solve the simplest case  of $N=2$.   

   The trivial solution is the matrix with $[X,Y] = 0$.  In this case we can choose 

$$ X  =\left[ \begin{array}{cc} 
 x_0      & 0\\
0 & \pm x_0\
\end {array}\right] , ~~~~~~~  Y  =\left[ \begin{array}{cc} 
 y_0      & 0\\
0 &  \pm y_0\
\end {array}\right] ,  \eqno{(3.1)}$$
\\
in which the coordinates $(x_0,y_0)$ and $(\pm x_0, \pm y_0)$ are the locations of the two $D_0$ branes respectively.   Substituting the above solution into (2.19) and (2.20) we find that $x_0^2 + y_0^2 $ is a complex number.   This means that the coordinates become unphysical complex values.   Therefore the original trivial state at $B=0$ become unstable under Melvin background and may be transfered into another physical and nontrivial stable state.  (Note that the trivial solution with $x_0 =y_0 =0$ at $B=0$ background will have infinite energy at finite $B$ background as can be seen from (2.18).)   In fact the potential form in (2.18) tells us that any trivial state, i.e. $[X,Y] =0$, at $B=0$ background will run away to infinite.  Thus  there does not exit any static trivial solution at Melvin background.

The static nontrivial solution we found is 

$$X=\pm ~c~{\sigma_x\over 2},~~~~~~Y = \pm ~c~{\sigma_y\over 2},$$
or
$$X=\pm ~c~{\sigma_y\over 2},~~~~~~Y = \pm ~c~{\sigma_x\over 2}\\
, \eqno{(3.2)}$$
\\
in which $c$ is a function of $B$ and can be easily determined from (2.20) and (2.21).  (For example, in the case of small magnetic field we have $c~ = \left(8 B^2\right)^{1\over 6}$.  Note that we have used the scale $P_-=- 1$.)  Substituting the above solutions into Hamiltonian (2.18) we see that this solution has a finite positive energy.    Thus we conclude that this configuration is a nontrivial solution in the Melvin background.  

   Note that  the authors in [10] had constructed the gravity solution for a fluxbrane expanded to sphere due to the extra dielectric effect.   The property is different from our finding.   

\section {Nonstatic Solutions of Melvin Matrix Model: Noncommutative Ring}
In this section we will try to find an nonstatic solution from the Hamiltonian (2.17).   We consider the following ansatz

$$X= x(t)~{\sigma_x} =  R cos(f(t)) ~{\sigma_x},~~~~~~Y = y(t)~{\sigma_y} = R sin(f(t)) ~ {\sigma_y},\eqno{(4.1)}$$
\\
in which $R$ is a constant value and  $f(t)$  a time-dependent function to be determined.  In this ansatz the Hamiltonian (2.17) becomes

$$H={\sqrt{2}\over \  (2+B^2)^{1/2}}{2\over \  (2+B^2)} \left[ (\dot x)^2+(\dot y)^2\right] + {-\sqrt{2}B^2\over \ 8 (2+B^2)^{3/2}}\left[\left(4B^2+B^4\right) R^2 - {\left(2+B^2\right)^2\over R^2}\right]~~~$$
$$+{1\over 8}{\sqrt {2} \left[4+(2+B^2)^2\right] \over \  (2+B^2)^{1/2}} x^2 y^2  =R^2~ \alpha (B^2) \left(\dot {f(t)}\right)^2 + R^4 ~ \beta (B^2) \left(sin(2f(t))\right)^2 + \gamma (R^2,B^2),\eqno(4.2)$$
\\
in which the detailed form of  the functions $ \alpha, \beta$ and  $\gamma $ are irrelevant to the following analyses.  

   We now begin to analyze the solution (4.1).  The equation (4.2) may be regarded as a particle moving under the potential $\sim  \left(sin(2f(t))\right)^2$.   As the Hamiltonian does not explicitely  dependent on time this is an energy-conservation system.  Thus, depending the initial energy, the particle may be oscillating or moving to infinite.  (Note that the coordinate in here is the function $f(t)$.)  As the radius $R^2$, (which is defined  by $x(t)^2+y(t)^2$), is a constant the solution therefore represents a ring which does not change the shape.   The ring is called a noncommutative ring as the coordinate $X$ and $Y$ in (4.1)  is a matrix which does not commutate to each other.   Also, as $x(t)$ and $y(t)$ is time dependent, the noncommutative ring will be rotating.   Depending on the function of $f(t)$, (which is  dependent of  the $B$ and $R$)  the ring may be rotating in righthand or in lefthand.   It may also be rotating form righthand to lefthand or form lefthand to righthand.

   Note that without the magnetic field the rotating membranes had been found by Harmark and Savvidy [17].

\section  {Discussion}\
   Let us finally discuss our results.

1. Historically, Dasgupta, Sheikh-Jabbari, and Raamsdonk [15] had successfully derived the BMN matrix model (which is proposed by Berenstein, Maldacena, and  Nastase [16] to describe M-theory on the maximally supersymmetric pp-wave)  directly as a discretized theory of  supermembranes in the pp-wave background.   In this paper, along this line,  we have applied the method of the regularized theory of light-cone gauge memberbranes [13] to find the matrix model in the Kaluza-Klein Melvin background.   

2.  Although the closed form of the Hamiltonian (2.17) is found it is difficult to use it to solve the problem.   We therfore study the simplest case of two $D_0$-branes in the static system.   We have found that the brane in the Melvin background have a solution which is a nontrivial configuration.   We belive that this is a general property for any value of $N$ and twist parameter $B$.  

3.  The Hamiltonian (2.18) has two mass terms , $ \sim B^2(X^2+Y^2)$.  However, these terms could not induce a nontrivial fuzzy sphere solution.  In fact, to find the fuzzy sphere solution we need the dielectric effect or, at least, the three mass terms, $ \sim B^2(X^2+Y^2+Z^2)$ [18].  

4.  It is reasonable to conjecture that the solution (3.2) is just one of the stationary state in the rotating solution of (4.1).   Our solution has shown that two $D_0$  propagating in the magnetic tube field background may expand into a rotating noncommutative ring.  We belive also that this is a general property for any value of $N$ and twist parameter $B$.    Note that the rotating behavior described in $f(t)$  is dependent of the ring radius $R$ and magnetic field $B$.   The futhermore physical properties of the nontrivial solution (4.1), such as its stability, the radiation during rotating, ...etc. [17], is remained to be investigated. 

5.  We have in this paper considered only the Boson Hamiltonian.   The Fermion part can be obtained by the supersymmetrization.    The interaction between a pair of the fluxbranes,  the recombination of the intersection fluxbranes and other dynamics of the fluxbranes can be investigated from the complete Hamiltonian in the Melvin matrix model.

   Work on these problems are in progress.

\newpage
\begin{center} {\large \bf  References} \end{center}
\begin{enumerate}
\item M.A. Melvin, ``Pure magnetic and electric geons,'' Phys. Lett. 8 (1964) 65.
\item G.~W.~Gibbons and D.~L.~Wiltshire, ``Space-time as a membrane in higher dimensions,'' Nucl.\ Phys.\ B287 (1987) 717 [hep-th/0109093]; \\G.~W.~Gibbons and K.~Maeda, ``Black holes and membranes in higher dimensional theories with dilaton fields,'' Nucl.\ Phys.\ B298 (1988) 741.
\item  J.~G.~Russo and A.~A.~Tseytlin, ``Exactly solvable string models of curved space-time backgrounds,'' Nucl.\ Phys.\ B449 (1995) 91 [hep-th/9502038];``Magnetic flux tube models in superstring theory,'' Nucl.\ Phys.\ B461 (1996) 131 [hep-th/9508068].
\item F.~Dowker, J.~P.~Gauntlett, D.~A.~Kastor and J.~Traschen, ``The decay of magnetic fields in Kaluza-Klein theory,'' Phys.\ Rev.\ D52 (1995) 6929 [hep-th/9507143].
\item M.~S.~Costa and M.~Gutperle, ``The Kaluza-Klein Melvin solution in M-theory,'' JHEP 0103 (2001) 027 [hep-th/0012072].
\item  A. Adams, J. Polchinski and E. Silverstein,  ``Don't Panic! Closed String Tachyons in ALE Spacetimes,''  JHEP 0110 (2001) 029 [hep-th/0108075].
\item J.~R.~David, M.~Gutperle, M.~Headrick and S.~Minwalla, ``Closed string tachyon condensation on twisted circles,''  JHEP 0202 (2002) 041 [hep-th/0111212].
\item T.~Takayanagi and T.~Uesugi, ``Orbifolds as Melvin geometry,''   JHEP 0112 (2001) 004 [hep-th/0110099];\\
S. Minwalla and  T. Takayanagi,  ``Evolution of D-branes Under Closed String Tachyon Condensation,''  JHEP 0309 (2003) 011 [hep-th/0307248]
\item M.~Gutperle and A.~Strominger, ``Fluxbranes in string theory,''
JHEP 0106 (2001) 035 [hep-th/0104136]; \\R.~Emparan and M.~Gutperle, ``From p-branes to fluxbranes and back,'' JHEP 0112 (2001) 023 [hep-th/0111177].
\item  M.~S.~Costa, C.~A.~Herdeiro and L.~Cornalba, ``Flux-branes and the dielectric effect in string theory,'' Nucl.\ Phys.\ B619 (2001) 155. [hep-th/0105023];\\ T.~Takayanagi and T.~Uesugi,  ``D-branes in Melvin background,'' JHEP 0111 (2001) 036 [hep-th/0110200].
\item  L. Motl, ``Melvin Matrix Models,'' [hep-th/0107002]
\item Y. Sekino and  T. Yoneya, ``From Supermembrane to Matrix String    Nucl.Phys. B619 (2001) 22-50,''  [hep-th/0108176]
\item B. de Wit, K. Peeters, J. Plefka, ``Superspace Geometry for Supermembrane  Backgrounds,''  Nucl.Phys. B532 (1998) 99, [hep-th/9803209].
\item  T.~Banks, W.~Fischler, S.~H.~Shenker and L.~Susskind,
``M theory as a matrix model: A conjecture,'' Phys.\ Rev.\ D 55 (1997) 5112
[hep-th/9610043].
\item  D.~Berenstein, J.~Maldacena and H.~Nastase, ``Strings in flat space and pp waves from  N= 4  Super Yang Mills'', JHEP  0204 (2002) 013,  [hep-th/0202021].
\item K.~Dasgupta, M.M.~Sheikh-Jabbari and M.~Van Raamsdonk, 
``Matrix Perturbation Theory For M-Theory On a PP-Wave'', JHEP  0205 (2002) [hep-th/02002021].
\item T. Harmark and K. G. Savvidy ,``Ramond-Ramond- Field Radiation from Rotating Ellipsoidal Membranes,''  Nucl.Phys. B585 (2000) 567, [hep-th/0002157].
 \item R.C. Myers, ``Dielectric-Branes'',  JHEP 9912 (1999) 022 [hep-th/0205185].

\end{enumerate}
\end{document}